\documentclass{egpubl}
\usepackage{egTechReport2019}

\ConferenceSubmission   

\usepackage{booktabs} 
\usepackage[T1]{fontenc}
\usepackage{dfadobe}  

\usepackage{cite}  
\BibtexOrBiblatex
\electronicVersion
\PrintedOrElectronic
\usepackage{egweblnk} 

\usepackage{algorithm}
\usepackage{algpseudocode}
\usepackage{adjustbox}
\usepackage{xspace}
\usepackage{relsize}

\usepackage[dvipsnames]{xcolor}
\usepackage{amsmath}
\usepackage{listings}
\usepackage{amssymb}
\usepackage{subcaption}
\usepackage{tabularx}
\usepackage{xcolor}
\usepackage{pgfplots}
\usepackage{tikz}
\usepackage{tabu}
\usepackage{mdframed}

\newmdenv[leftline=false,rightline=false,innertopmargin=+0.4cm]{topbot}

\definecolor{dblue}{rgb}{0,0,0.4}
\definecolor{bblue}{HTML}{4F81BD}
\definecolor{dred}{rgb}{0.4,0,0}
\definecolor{rred}{HTML}{C0504D}
\definecolor{dgreen}{rgb}{0,0.4,0}
\definecolor{ggreen}{HTML}{9BBB59}
\definecolor{dpurple}{rgb}{0.4,0,0.4}
\definecolor{ppurple}{HTML}{9F4C7C}
\definecolor{dyellow}{rgb}{0.4,0.4,0}
\definecolor{yyellow}{rgb}{1,1,0}
\definecolor{dlime}{rgb}{0.1,0.4,0}
\definecolor{llime}{rgb}{0.749,1,0}
\definecolor{dkingblue}{rgb}{0,0.1,0.6}
\definecolor{kkingblue}{rgb}{0,0.2,1}
\definecolor{dgray}{rgb}{0.3,0.3,0.5}
\definecolor{ggray}{rgb}{0.75,0.75,0.85}
\definecolor{dblack}{rgb}{0,0,0}
\definecolor{bblack}{rgb}{0.1,0.1,0.1}

\DeclareCaptionSubType*{figure}

\definecolor{dkgreen}{rgb}{0,0.6,0}
\definecolor{dkblue}{rgb}{0,0,0.6}
\definecolor{gray}{rgb}{0.5,0.5,0.5}
\definecolor{mauve}{rgb}{0.58,0,0.82}

\definecolor{commentgreen}{RGB}{2,112,10}
\definecolor{eminence}{RGB}{108,48,130}
\definecolor{weborange}{RGB}{255,165,0}
\definecolor{frenchplum}{RGB}{129,20,83}

\newcommand{\Test}[1]{\expandafter\hat#1}

\title[Augmenting Image Warping-Based Remote Volume Rendering with Ray Tracing]{
Augmenting Image Warping-Based Remote Volume Rendering with Ray Tracing}

\author[Zellmann]{\parbox{\textwidth}{\centering
    Stefan Zellmann
    \thanks{zellmann@uni-koeln.de, Department of Computer Science, University of
    Cologne}\orcid{0000-0003-2880-9090}
    \quad
  }
}

\begin{document}

\teaser{ \centering \resizebox{0.98\textwidth}{!}{
\includegraphics[width=0.33\textwidth]{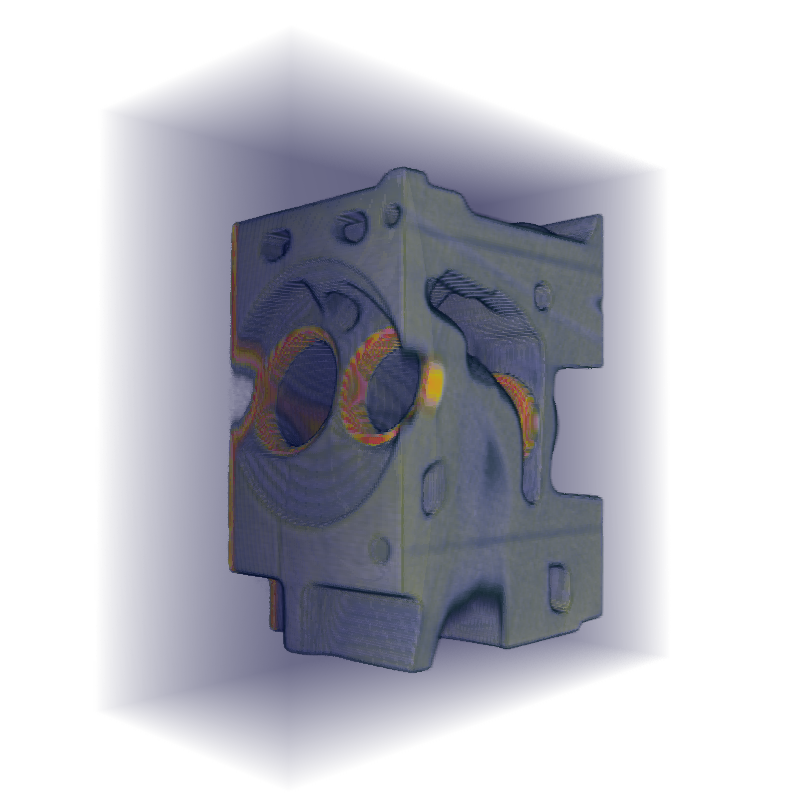}
\includegraphics[width=0.33\textwidth]{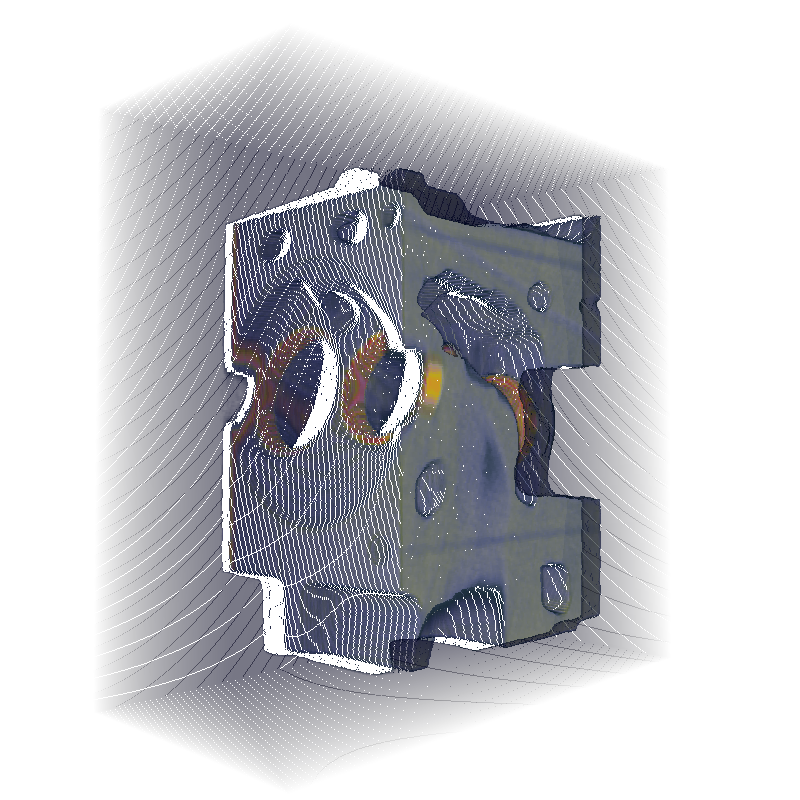}
\includegraphics[width=0.33\textwidth]{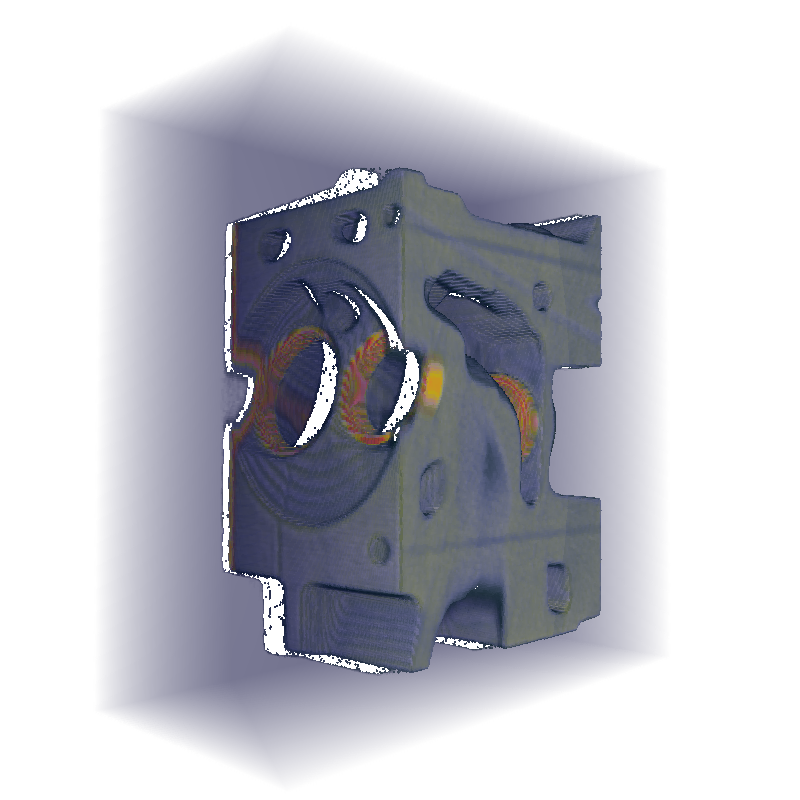}
 }
\caption{\label{fig:teaser}
Image warping-based remote rendering for volume data that reprojects 2.5D depth
images from outdated frames to hide latency. Left: reference image. Middle: 2.5D
image rendered as OpenGL point primitives. Right: the same reprojected 2.5D
data, but rendered as an object-space point cloud using ray tracing. The ray
tracing-based technique is less susceptible to reprojection artifacts as the
footprint of the reprojected points depends on the distance to the viewer.
} }

\maketitle

\begin{abstract}
We propose an image warping-based remote rendering technique for volumes that
decouples the rendering and display phases. Our work builds on prior work that
samples the volume on the client using ray casting and reconstructs a z-value
based on some heuristic. The color and depth buffer are then sent to the client
that reuses this depth image as a stand-in for subsequent frames by warping it
according to the current camera position until new data was received from the
server. We augment that method by implementing the client renderer using ray
tracing. By representing the pixel contributions as spheres, this allows us to
effectively vary their footprint based on the distance to the viewer, which we
find to give better results than point-based rasterization when applied to
volumetric data sets.
\end{abstract}

\section{Introduction}
Remote rendering is an important technique to overcome the typical bandwidth
limitations in in-situ scenarios, or when accessing graphics workstations over
LAN or WAN using thin clients. Remote rendering algorithms can be classified by
the type of data---image pixels, proxy geometry, etc.---that is sent over the
network, and by the amount of post-processing that needs to be done on the
client, with the spectrum ranging from \emph{send-image} over
\emph{send-geometry} to \emph{send-data} approaches~\cite{bethel:2012}.
According to this classification, send-image implementations execute the full
rendering pipeline on the remote server or workstation, while the client is
responsible only for display.

We present a remote rendering technique based on prior work by Zellmann
et al.~\cite{zellmann:2012} that decouples the rendering and display phases. By
that, latency introduced by the network or the rendering algorithm itself can be
hidden and the user interface always remains responsive. That is an important
property for certain usage scenarios---like e.g.\ virtual reality with head
tracking---and can help to improve the overall user experience.

With real-time ray tracing nowadays being widely available even on consumer
hardware, we present and evaluate a simple improvement to the algorithm by
Zellmann et al. that does not render images directly, but interprets image
pixels as object space splats and renders them as transparent spheres in a ray
tracer. This enhancement can help to conceal reprojection artifacts and is
specifically helpful for remote \emph{volume} rendering, which the algorithm was
originally designed for.

The paper is structured as follows. In Section~\ref{sec:background} we briefly
summarize the algorithm by Zellmann et al. In Section~\ref{sec:related} we
present related work regarding remote rendering, and in Section~\ref{sec:method}
we present our enhancements to the remote rendering algorithm. We present
results in Section~\ref{sec:results}, discuss those in
Section~\ref{sec:discussion}, and conclude the publication in Section
\ref{sec:conclusion}.

\section{Background}
\label{sec:background}
\label{sec:related}
\begin{figure*}[tb]
\centering
\includegraphics[width=\textwidth]{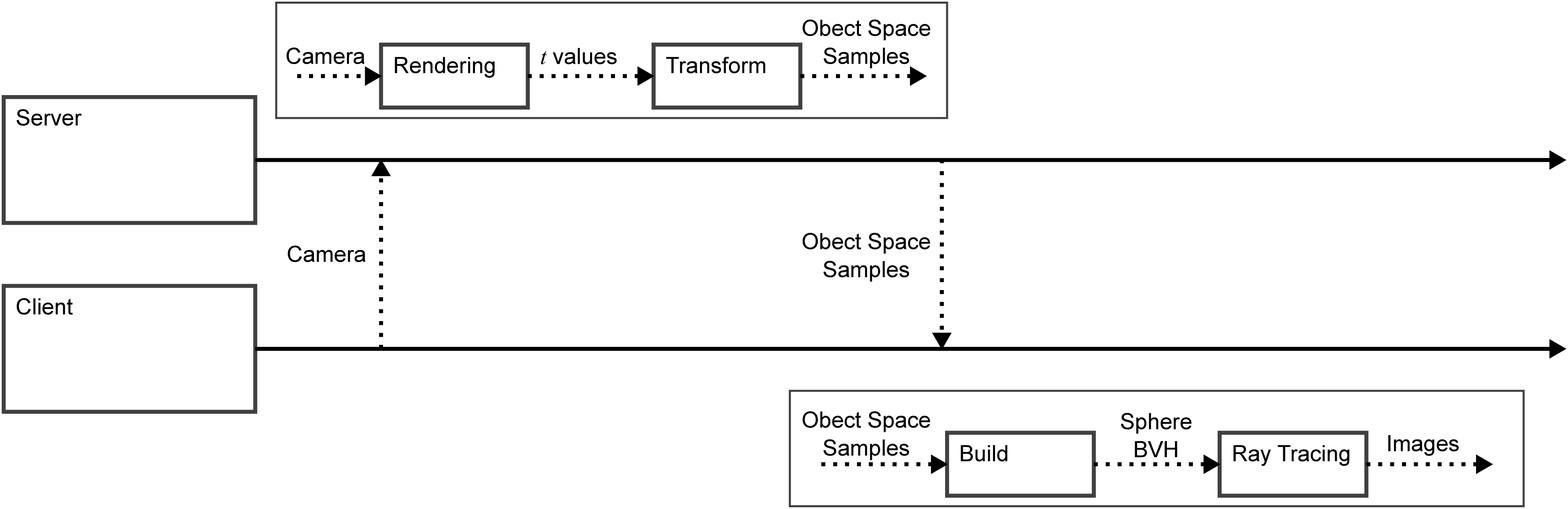}
\caption{\label{fig:overview}
Overview of the client / server architecture for remote rendering. After
receiving camera information, the server employs ray marching to generate
primary ray parameter values $t_i$ associated with each pixel and transforms
those into object space samples afterwards. The client receives those object
space samples, reinterprets them as colored, semi-transparent spheres, builds a
BVH over these and later ray traces them using a simple first-hit ray tracing
query and with no shading other than the constant sample color.
\vspace{-1em}
}
\end{figure*}
We present a volume remote rendering technique that broadly falls into the range
of \emph{send-image} approaches~\cite{bethel:2012}, where fully rendered images
are sent over the network. Our technique builds on prior work
by Zellmann et al.~\cite{zellmann:2012}. In that work, the authors decoupled
the rendering and display phases by displaying final images at a different rate
than delivered by the server. Whenever the camera position changes, the current
pixel buffer is warped according to the new transformation; whenever the server
sends an updated image, the pixel buffer is updated. When the rendering phase on
the server and the display phase on the client operate at exactly the same rate,
that will result in the client always displaying the correct image. With
different rates, the client will display pixels from outdated frames that thus
appear warped. That effect is more exaggerated as the the delay between the two
phases grows bigger. On the other hand, assuming that an image can be displayed
faster than it can be rendered, the user interaction appears smooth because user
input is processed at high rates.

In order to improve the impression of depth of the warped images, Zellmann et
al. augment the 2D images with an additional depth buffer. Instead of warping
flat images, the authors render the image pixels as point primitives on the GPU.
They therefore use OpenGL vertex buffer objects and color textures that are
updated when the server sends a new image.

A challenge specific to volume rendering or rendering of transparent objects in
general is that there is no definitive depth value that represents all the
surfaces contributing to the pixel. With volume rendering, there are not even
definitive surfaces to choose from to contribute such a depth value. Zellmann et
al. therefore evaluated a number of heuristics that assume that the server
performs one or multiple ray marching passes over the volume data. The results
the authors presented were mixed, with certain heuristics being more effective
depending on volume data and transfer function. The method works best when the
transfer function contains high-frequency iso values that are mapped to
relatively high opacity.

The most effective heuristic to determine depth values is still open to
debate, and more recent work has presented approaches and heuristics that are
arguably more effective for volume rendering~\cite{lochmann:2016,martin:2020}.
Still, the warping technique by Zellmann et al.
is fast and thus a simple and easy to implement improvement to achieve better
interactivity over techniques where rendering and display phase are locked.

\section{Related Work}
Send-image remote rendering is a popular approach that has for example been
proposed by Stegmaier et al.~\cite{stegmaier:2002}.  Visualization tools like
ParaView~\cite{ahrens:2005} support client / server rendering modes that exchange
images; also, dedicated remote rendering tools like VirtualGL
\cite{virtualgl:2020} allow the user to use accelerated graphics over the
network via send-image remot rendering. We refer the reader to the text book on
high performance visualization edited by Bethel et al.~\cite{bethel:2012} and
the survey article by Shi and Hsu~\cite{shi:2015} for an introduction to and a
good general overview of the several remote rendering techniques.

The idea to use image warping in low bandwidth scenarios is relatively old and
was e.g.\ proposed by Bao et al.~\cite{bao:2003}. Research has focused on
augmenting mere send-image approaches with additional data like depth
buffers~\cite{zellmann:2012}, image layers~\cite{lochmann:2016}, or even light
fields~\cite{martin:2020}. The work by Shi et al.~\cite{shi:2012} has focused on
image warping techniques using depth images targeting mobile devices. The paper
by Pajak et al.~\cite{pajak:2011} has explored compression techniques based on
spatio-temporal upsampling on the client that also includes use of depth buffers
sent over network.

The work by Schied et al.~\cite{schied:2017,schied:2018} on spatio-temporal
variance-guided filtering also goes roughly in the same direction as our approach
as it is based on rendering with outdated data, but their technique is based
on sample accumulation and on extrapolating samples into the future using motion
vectors, whereas our approach, in comparison, predicts the present image samples
based on past image data.

\section{Method}
\label{sec:method}
We propose to augment the remote rendering algorithm by Zellmann et al.
\cite{zellmann:2012} by switching from an image-based approach to an object order
approach based on real-time ray tracing. Instead of generating 2.5D data on the
remote server, we generate world space samples carrying the final composited
color from volume ray marching along. On the client, we transform those samples
to a point cloud that we render using first-hit ray tracing. We also
experimented with multi-hit ray tracing~\cite{zellmann:2017}, but as the results
regarding image quality were mixed, we stuck with this simple approach that just
colorizes each sphere according to its designated pixel color.

A challenge of the warping technique with rasterization is that the point
primitives' size is fixed to a certain number of pixels, and this causes a
variety of artifacts that can be avoided when the 2.5D data set is represented
with solid objects. In the latter case, solids that are closer to the viewer
will cover more pixels. Effectively, this can be regarded as splatting, where
the footprint of the splats decreases with increasing distance to the viewer.
While rendering of geometrically complex objects like tessellated spheres with
OpenGL is prohibitive memory-wise, with real-time ray tracing and arbitrary user
geometry it is a viable option to render the 2.5D geometry as spheres that are
represented as real quadrics. With this extension, we hope to reduce the impact
of the artifacts encountered with rasterization-based 2.5D image
warping---especially in the presence of volumetric data and semi-transparent
pixels.

\subsection{Object space samples}
On the server side, we render the volume using a ray marcher that employs one of
the heuristics presented by Zellmann et al. to estimate which depth value
represents the volume best. We march rays $r = o + \vec{d}t$ with origin $o$,
direction vector $\vec{d}$ and ray parameter $t$. When a representative depth
according to one of the heuristics was found, that depth is associated with a
certain value for $t=t_i$. Zellmann et al. first compute an object space coordinate
for that value:
\begin{equation}
p = o + \vec{d}t_i
\label{eq:point}
\end{equation}
and then \emph{reproject} that to obtain a position in OpenGL window coordinates
by applying the viewing and camera transforms as well as viewport transform.

In contrast to that, we directly send object space samples to the client that
are comprised of the coordinates from Equation~\ref{eq:point} and a footprint
radius. We quantize the object space coordinate and pack it together with the
radius into 64 bits. The ray marcher fills a screen-sized buffer with object
space samples and associated volume-rendered colors. We currently just set the
footprint radius to half the size of a voxel's diagonal when the ray hit the
volume and integrated a color with non-zero opacity. Otherwise, we set the
radius to zero. Before sending we compact that buffer by moving all the samples
with zero radius to one side and send only those object space samples with
non-zero radius.

On the client, when we received buffers with object space samples and colors, we
reinterpret the object space samples as semi-transparent spheres, build a
bounding volume hierarchy from those using the LBVH algorithm
\cite{lauterbach:2009,zellmann:2019}, and render them as a point cloud using ray
tracing. The overall process is depicted in in Figure \ref{fig:overview}.

\subsection{Implementation}
We implemented the framework described above using NVIDIA CUDA and network
communication using the C++ Boost Asio library. We use a standard ray marching
volume renderer that writes out colors and $t$ values to an off-screen buffer.
When the client sends an updated camera, the server performs volume rendering,
fills the off-screen buffers and sends them to the client. Camera motion
processing and request dispatch are handled in threads separate from the main
rendering thread to decouple this from the display phase. After rendering, the
server transforms and quantizes the $t$ values and sends both position and color
buffers to the client, along with the camera the buffers are associated with so
the client does not need to keep track of this. The client upon receiving the
data will build an LBVH over the positions that is subsequently used to render
the point cloud.
\vspace{-1em}
\section{Results}
\label{sec:results}
\begin{figure*}[tb]
    \includegraphics[width=0.195\textwidth]{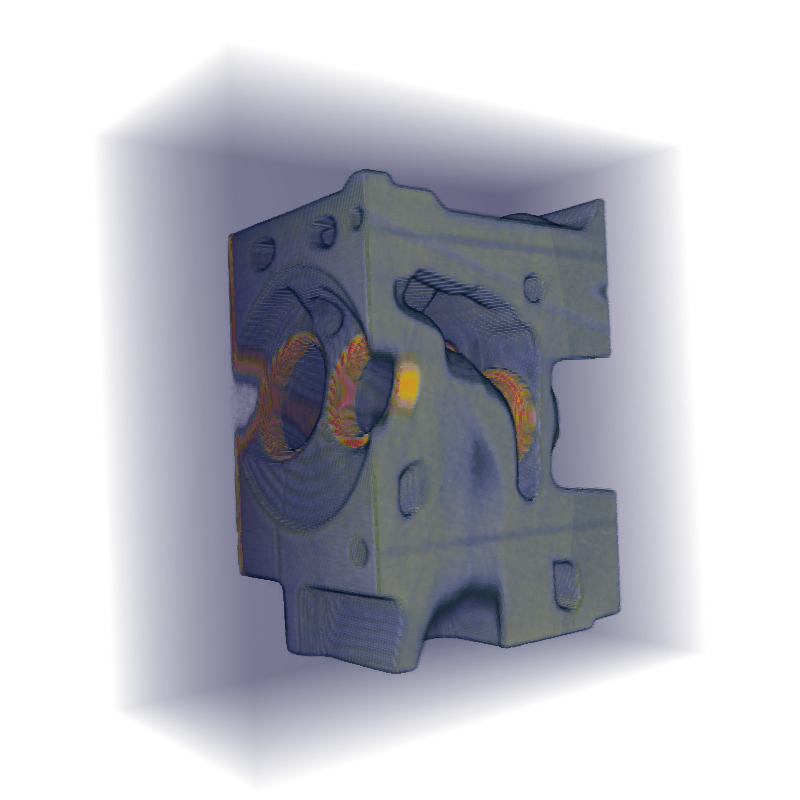}
    \includegraphics[width=0.195\textwidth]{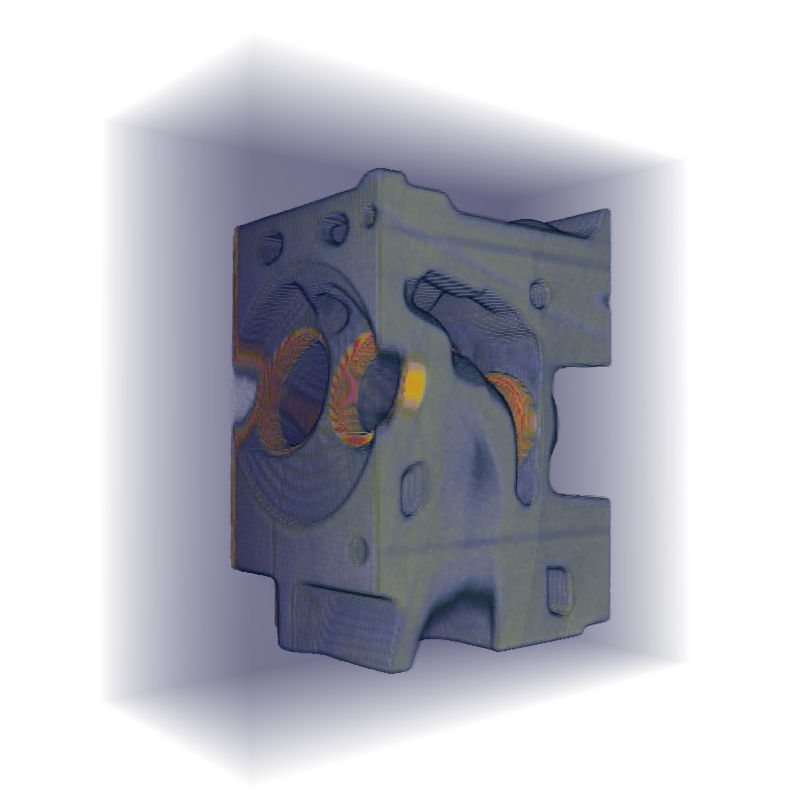}
    \includegraphics[width=0.195\textwidth]{cam3}
    \includegraphics[width=0.195\textwidth]{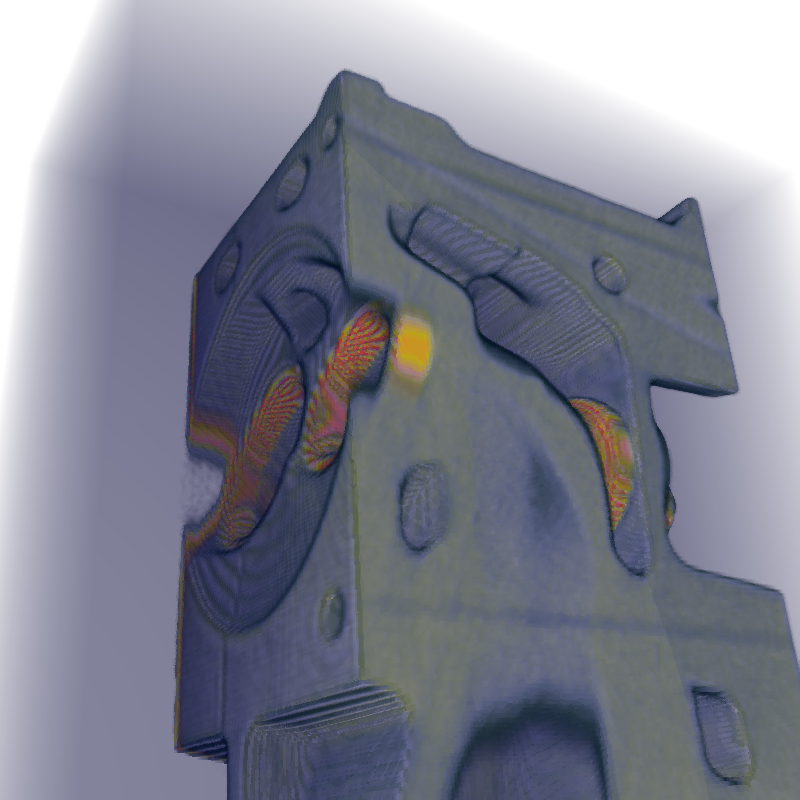}
    \includegraphics[width=0.195\textwidth]{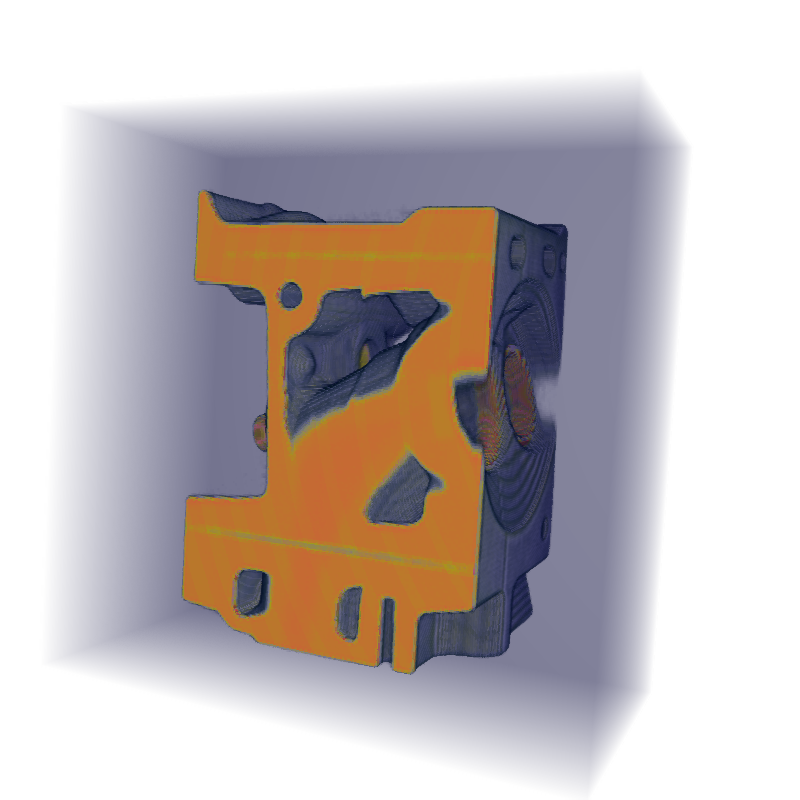}\\
    \includegraphics[width=0.195\textwidth]{cam1}
    \includegraphics[width=0.195\textwidth]{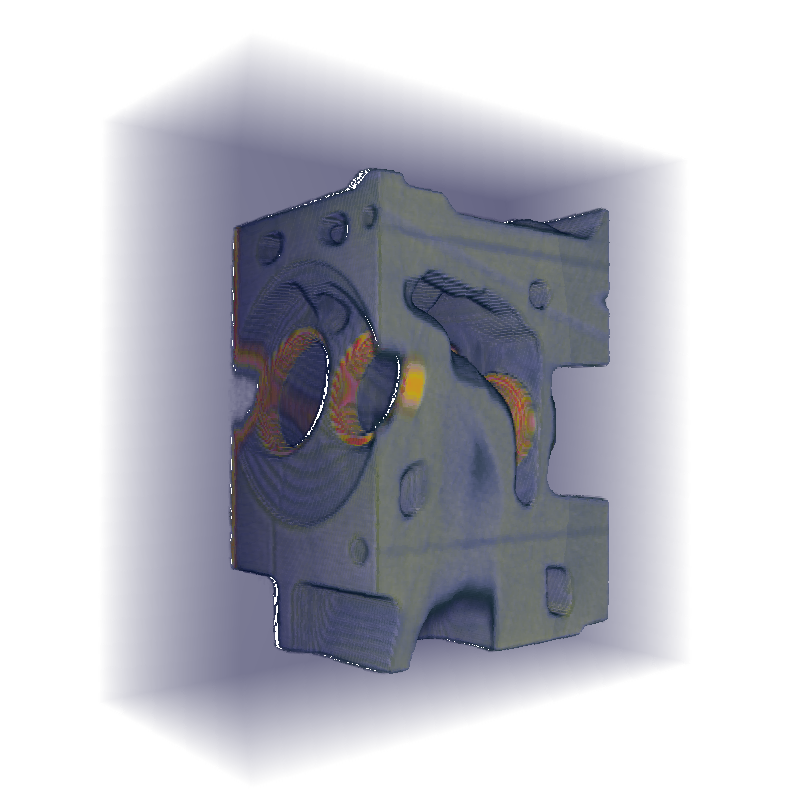}
    \includegraphics[width=0.195\textwidth]{cam3_warped}
    \includegraphics[width=0.195\textwidth]{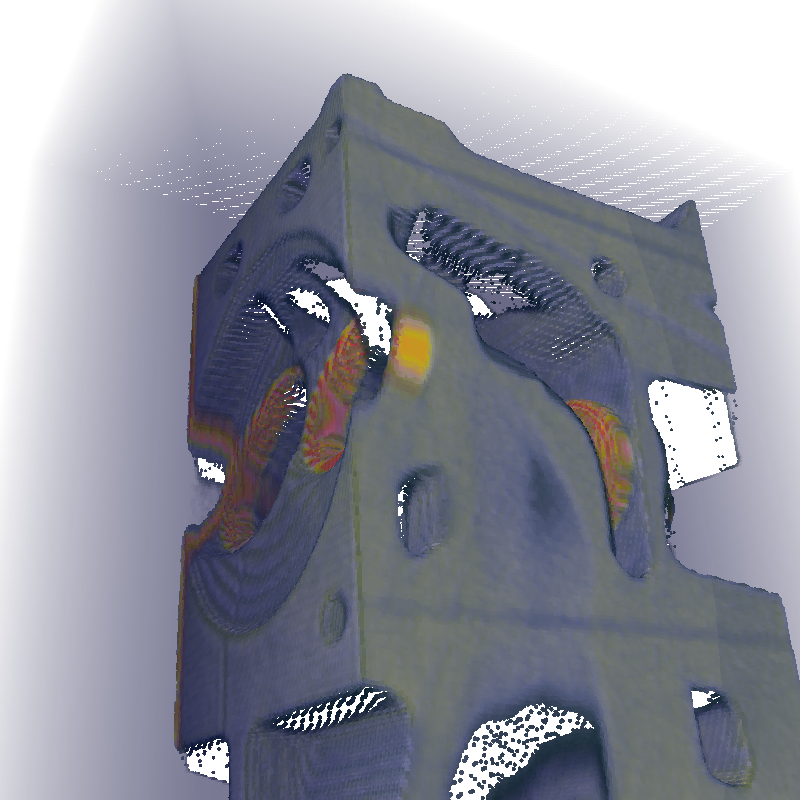}
    \includegraphics[width=0.195\textwidth]{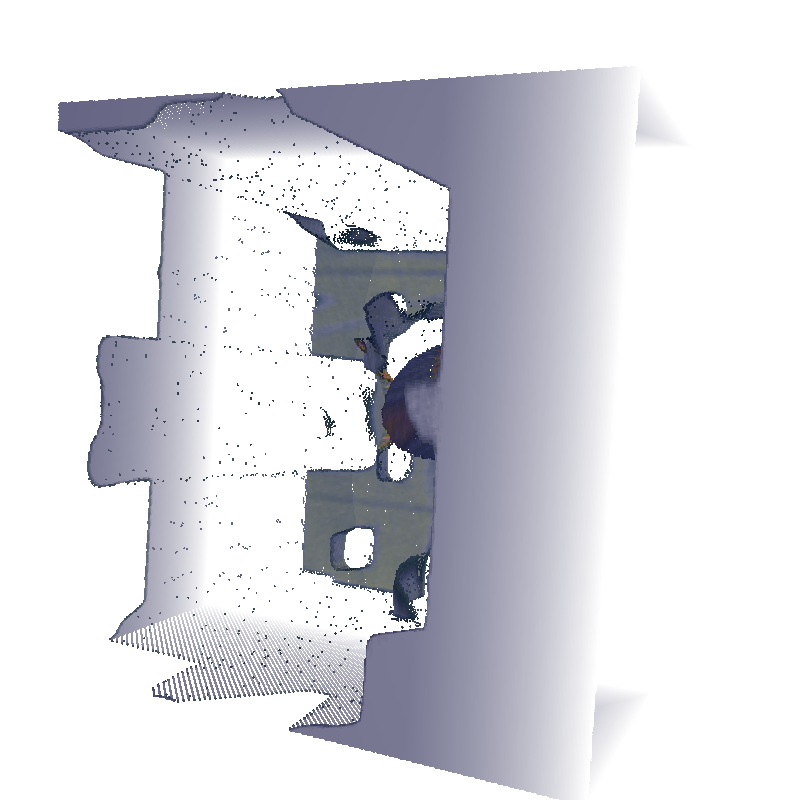}\\
    \includegraphics[width=0.195\textwidth]{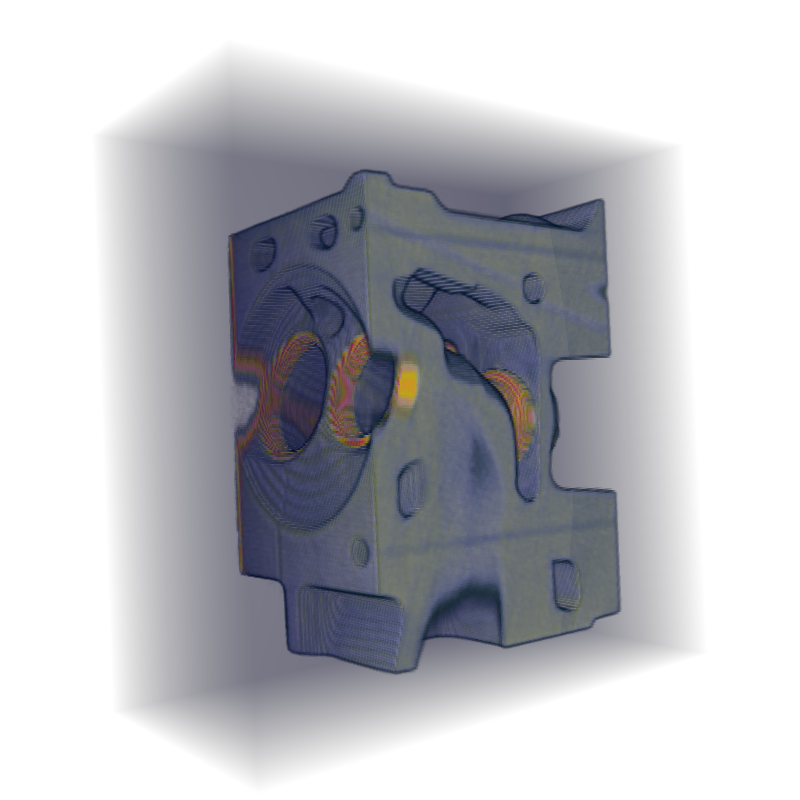}
    \includegraphics[width=0.195\textwidth]{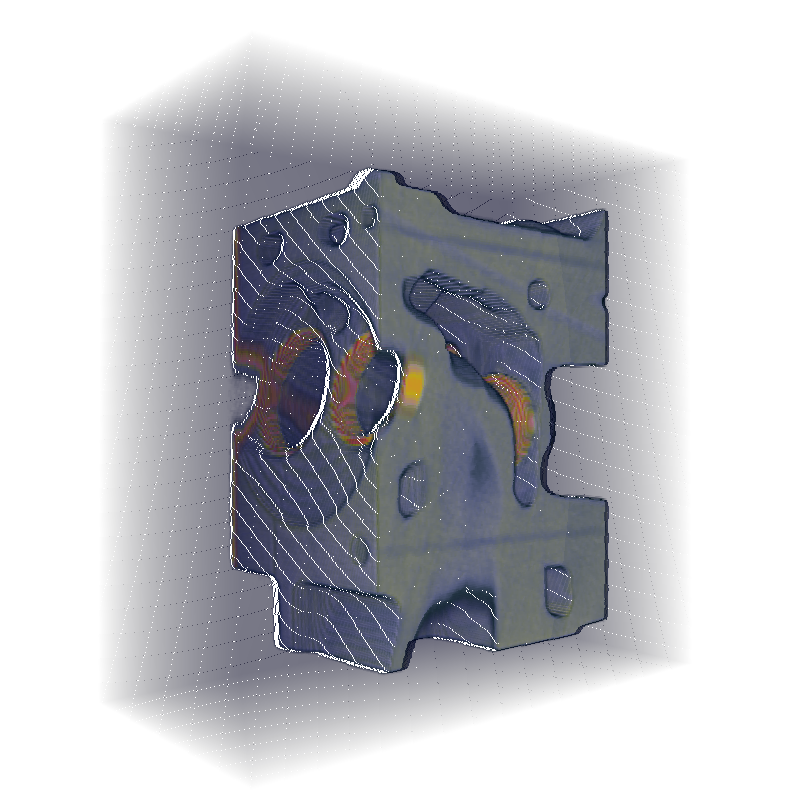}
    \includegraphics[width=0.195\textwidth]{cam3_gl}
    \includegraphics[width=0.195\textwidth]{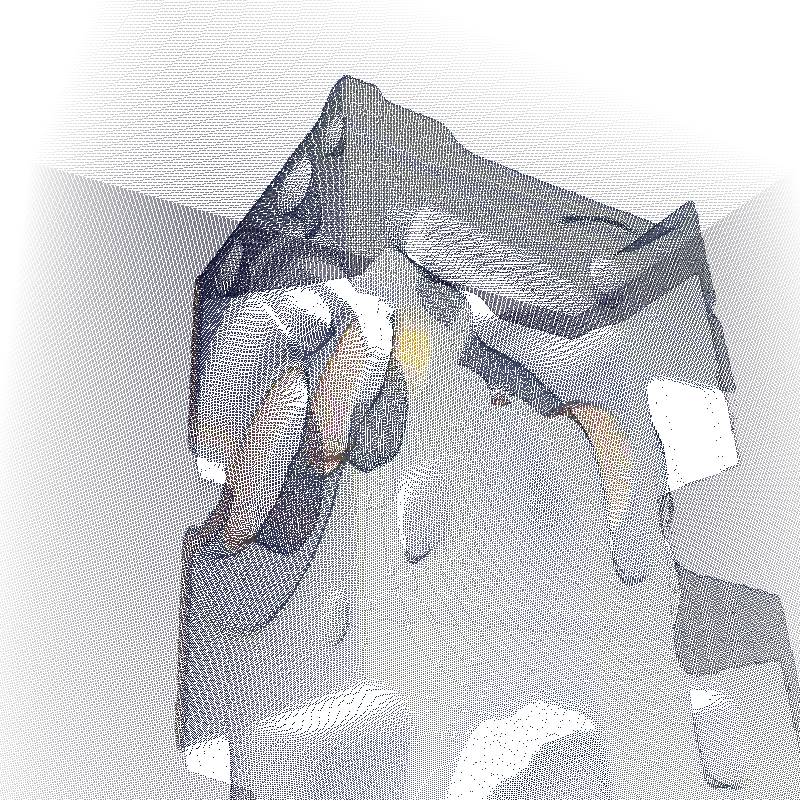}
    \includegraphics[width=0.195\textwidth]{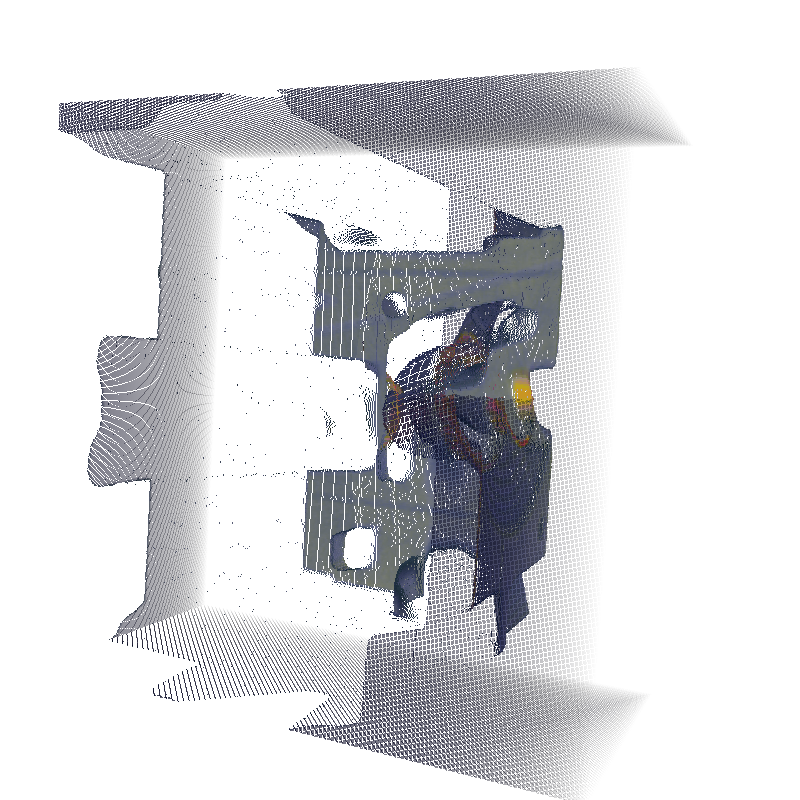}
\caption{
\label{fig:comparison}
Comparison between OpenGL reprojection with rasterized points, and our ray
tracing technique rendering object space samples as spheres. Top row: a sequence
of successive frames obtained using volume rendering, with the camera position
gradually changing to positions that show the volume from different viewing
angles. Middle row: the same sequence of frames; the left image is subsequently
warped as a 2.5D point cloud to the viewing positions from the top row using
our ray tracing technique. Bottom row: the sequence of frames is warped in the
same manner, but with OpenGL point rasterization. Rendering artifacts especially
in regions that appear volumetric and transparent are much more exaggerated even
when the camera is only slightly moved.
\vspace{-1em}
}
\end{figure*}
In order to compare the effectiveness of our approach quality-wise, we
implemented the reference method by Zellmann et al. using OpenGL point
rendering. Figure~\ref{fig:comparison} shows this qualitative comparison. In the
example a sequence of warped frames is obtained using the gradient heuristic.

\begin{table}[tb]
  \centering
  \begin{tabular}{|r|c|c|c|c|c|}
    \toprule
    ~ & View~1  & View~2    & View~3    & View~4    & View~5    \\
    \midrule
    Compact & \multicolumn{5}{c|}{$23$~ms} \\
    \midrule
    LBVH & \multicolumn{5}{c|}{$11$~ms} \\
    \midrule
    FPS  & 292 & 276  & 283  &  229  & 408  \\
    \bottomrule
  \end{tabular}
\caption{
Performance results for rendering the five views from Figure~\ref{fig:comparison}
with our ray tracing technique. We render with a viewport of $1024 \times 1024$
pixels. The point cloud after compaction is comprised of $623,584$ object space
samples.
\label{tab:performance}
\vspace{-2em}
}
\end{table}
For a performance evaluation we report compaction rates on the server (currently
using a serial implementation), LBVH construction time, as well as average
rendering performance for the five views from Figure~\ref{fig:comparison} in
Table~\ref{tab:performance}. We measured those values using an NVIDIA RTX~2080
GPU.

\section{Discussion}
\label{sec:discussion}
We find this simple extension to the original algorithm by Zellmann et al. to be
effective. As can be seen from Figure~\ref{fig:comparison}, the visual quality
when rendering the 2.5D point cloud as object space splats instead of OpenGL points
with a fixed size in pixels improves dramatically. We also
experimented with setting a variable point size in OpenGL mode but found this
setting to be hard to control as it biases the rendered results because the
opacity of the composited point sprites increases with an increase in fillrate.
Besides, depth compositing for point primitives with OpenGL is not performed per
fragment but per vertex; alternatives to point sprite rendering that would
mitigate these problems would be to use geometry shaders to expand the point
geometry into more complex geometric shapes.

We however found the rendering performance of the ray tracing method---this is
a software implementation based on NVIDIA CUDA---to be so high even though it
does not use RT Cores, that we deem ray tracing to be the most elegant solution
that with ray tracing-capable hardware is available on every 
contemporary commodity GPU. The overall pipeline is currently bound by
compaction performance on the server, which would however be easy to fix, e.g.,
by using the \texttt{remove\_if} standard algorithm from the C++ / GPGPU library
\texttt{thrust}. We deliberately have not reported network-related performance as that
would relate to the available bandwidth and latency of an individual network
connection. We currently send 32 bits per color and 64 bits per object space
sample (position and radius). The above images ($1024 \times 1024$ pixels),
after compaction, consist of $623,584$ individual object space samples.

\section{Conclusion and Future Work}
\label{sec:conclusion}
We presented a simple yet effective extension to the algorithm by Zellmann et al.
that massively improves the rendering artifacts that this remote volume
rendering technique otherwise suffers from by replacing the 2.5D point rendering
phase with an object space ray tracing phase. This potentially comes at moderate
additional costs regarding memory bandwidth as we have to store object space
coordinates, although we in turn can benefit from compaction on the server side.

Switching to a ray tracing pipeline presents us with a wealth of possibilities
that we intend to explore in the future and that this work lays the groundwork
for. Possible extensions are for example accumulation of samples across a couple
of frames or varying the radius of the object space samples according to their
opacity or according to uncertainty regarding the depth value we obtained:
pixels where we are uncertain where to place them along the viewing ray might be
smeared out across some interval that is longer the more uncertain we are.
Another interesting route to explore is compression of object space samples,
which would be important to obtain high throughput. Yet another extension to our
method would be hardware ray tracing using NVIDIA's RT cores, although the frame
rates we observe on the client with GPU software ray tracing are already very
high.

\footnotesize
\bibliographystyle{eg-alpha}
\bibliography{egbibsample}

\end{document}